\documentclass{ws-procs9x6}
\usepackage{graphicx}
\begin{document}

\title{Properties of Thermal Glueballs\footnote{\uppercase{T}he numerical calculation has been done on \uppercase{NEC SX-5} at \uppercase{O}saka \uppercase{U}niveristy.}
}

\author{NORIYOSHI ISHII}
\address{
  Radiation Laboratory,\\
  RIKEN (The Institute of Physical and Chemical Research),\\
  2-1 Hirosawa, Wako, Saitama 351--0198, Japan\\
  E-mail: ishii@rarfaxp.riken.go.jp}

\author{HIDEO SUGANUMA}
\address{
  Faculty of Science, Tokyo Institute of Technology,\\
  2--12-1 Ohokayama, meguro, Tokyo 152--8550, Japan
}



\maketitle

\abstracts{
We study the properties of the $0^{++}$ glueball at finite temperature
using SU(3) quenched lattice QCD.
We find a significant thermal effects near $T_c$.
We perform  the $\chi^2$ fit  analyses adopting two Ans\"atze  for the
spectral function,  i.e., the  conventional narrow-peak Ansatz  and an
advanced  Breit-Wigner Ansatz.   The  latter is  an  extension of  the
former,  taking account  of the  appearance  of the  thermal width  at
$T>0$.   We also perform  the MEM  analysis.  These  analyses indicate
that the thermal effect on the glueball is a significant thermal-width
broadening $\Gamma(T_c) \sim 300$ MeV together with a modest reduction
in the peak center $\Delta\omega_0(T_c) \sim 100$ MeV.
}

\newcommand{\Eq}[1]{Eq.~(\ref{#1})}
\newcommand{\Eqs}{Eqs.}
\newcommand{\Fig}[1]{Fig.~\ref{#1}}
\newcommand{\Bs}{\hspace*{-1em}}
\newcommand{\Hs}{\hspace*{1em}}

Near the  critical temperature $T_c$, the QCD  vacuum begins to change
its structure.  For instance,  the   string tension reduces and    the
spontaneously broken  chiral symmetry   begins to  restore  partially.
Since hadrons are composites of  quarks and gluons, whose  interaction
depends on the properties of the  QCD vacuum, it  is natural to expect
the structural changes of various hadrons.
In  this  direction,  there are  a  number  of  studies based  on  QCD
motivated  effective  models, which  predict  the  mass reductions  of
charmonium       and      light      $q\bar{q}$       mesons      near
$T_c$.\cite{hashimoto,hatsuda} In  BNL-RHIC experiments, these changes
are  considered  as  important  precritical  phenomena  of  QCD  phase
transition.  In  addition, lattice  QCD Monte Carlo  calculations have
been  performed to  examine the  pole-mass  shifts of  hadrons at  the
quenched level.\cite{taro,umeda}
%
%
The  $0^{++}$ glueball is  one of such hadrons,  which are expected to
exhibit structural changes near $T_c$.  Studies based on the effective
models\cite{ichie,ishii3}   suggest  a large  mass  reduction   of the
glueball as a natural consequence  of  the sizable difference  between
the $0^{++}$ glueball mass $m_{\rm  G}=1500-1700$ MeV and the critical
temperature $T_c = 260-280$ MeV.
In  this  paper, we  investigate  the  glueball  properties at  finite
temperature    by     means    of    lattice     QCD    Monte    Carlo
calculation.\cite{ishii,ishii2,ishii4}

We consider the temporal correlator $G(\tau) \equiv \langle \phi(\tau)
\phi(0) \rangle$  of the glueball operator  $\phi(\tau)$. The simplest
construction  of  the glueball  operator  is  the plaquette  operator,
which,  however, is known  to have  only a  negligible overlap  to the
lowest-lying $0^{++}$  glueball. To enhance the  overlap, the smearing
method  provides  us with  a  systematic  procedure  by extending  the
spatial  size  of the  operator.   In this  paper,  we  adopt the  APE
smearing.\cite{ishii2,ape}
The continuum expression of  the smeared glueball operator is obtained
in the Coulomb gauge as
\begin{equation}
  \phi
\propto
  \int {d^3y d^3z\over (2\pi)^{3/2} \rho^3}
  \exp\left( - {(\vec y - \vec z)^2 \over 2\rho^2} \right)
  G^a_{ij}(\vec y) G^a_{ij}(\vec z),
\label{smearing}
\end{equation}
where  $G^a_{ij}$ is   the    field strength   tensor,  and   $\rho\in
\mathbb{R}$ controls the size  of the spatial extension.  (For detail,
see Refs.~\refcite{ishii2}.)  By choosing $\rho$ appropriately, we can
maximize   the  contribution from   the   lowest-lying glueball in the
glueball correlator $G(\tau)$.

We  next  consider  the  spectral representation  of  $G(\tau)  \equiv
\langle \phi(\tau) \phi(0) \rangle$ as
\begin{equation}
  G(\tau)
=
  \int_{0}^{\infty} d\omega
  K(\tau,\omega) A(\omega),
\label{spect}
\end{equation}
with       $\beta\equiv       1/T$,       $K(\tau,\omega)       \equiv
{\cosh\left(\omega(\beta/2         -         \tau)\right)        \over
\sinh(\beta\omega/2)}$, and $A(\omega)$  denotes the spectral function
with its spatial momentum projected to zero.  At low temperature, peak
positions of $A(\omega)$  provide us with the pole  masses of hadrons.
Hence the  simplest parameterization  of $A(\omega)$ for  the $\chi^2$
fit analysis is given as\cite{ishii}
\begin{equation}
  A(\omega)
=
  C\left\{   \delta(\omega - m) - \delta(\omega+m)  \right\},
\label{narrow.peak.Ansatz}
\end{equation}
where $C$ and $m$ are used as the fit parameters, corresponding to the
strength and the  pole mass, respectively.  To adopt  this Ansatz, the
peak    should    be    narrow    enough.    We    will    refer    to
\Eq{narrow.peak.Ansatz}  as  the  ``{\em  narrow-peak  Ansatz}''.   At
$T>0$,  however, through  the interaction  with the  thermally excited
particles, the  thermal width is generated even  for stable particles.
As  $T$ gets  larger, the  effects of  the thermal  width  become more
important.
To  respect its  existence,  we adopt  the following  parameterization
as\cite{ishii2}
\begin{equation}
  A(\omega)
=
  C\left\{
    \delta_{\Gamma}(\omega - \omega_0) - \delta_{\Gamma}(\omega + \omega_0)
  \right\},
\label{breit.wigner.Ansatz}
\end{equation}
where                                   $\delta_{\Gamma}(\omega)\equiv
\frac1{\pi}\mbox{Im}\left(\frac1{\omega  -    i\Gamma}\right)$.   $C$,
$\omega_0$, and $\Gamma$ are used  as the fit parameters corresponding
to the strength, the peak center, and the thermal width, respectively.
We  will refer to \Eq{breit.wigner.Ansatz}  as the ``{\em Breit-Wigner
Ansatz}''.    Note     that   $\displaystyle       \lim_{\Gamma\to  0}
\delta_{\Gamma}(\omega)  =   \delta(\omega)$.  Hence, the Breit-Wigner
Ansatz is a natural extension of the narrow-peak Ansatz.

For these $\chi^2$ fit analyses, proper Ans\"atze for $A(\omega)$ have
to be  provided.  However,  in the very  vicinity of $T_c$,  and above
$T_c$, there may appear more complicated structures.
In this  sense, it  is desirable to  obtain $A(\omega)$  directly from
$G(\tau)$,  which,  however,  is  known  to be  an  ill-posed  problem
numerically.
Recently developed  Maximum Entropy Method (MEM)\cite{jarrell,asakawa}
can   deal  with   this   inverse  problem   adopting  the   following
Shannon-Jaynes entropy as
\begin{equation}
  S
\equiv
  \int_0^{\infty} d\omega
  \left[
    A(\omega) - m(\omega)
    - A(\omega)\log\left({A(\omega) \over m(\omega)}\right)
  \right],
\end{equation}
where $m(\omega)$ is  a real and positive  function referred to as the
default model function.     $m(\omega)$ is required to  reproduce  the
asymptotic behavior of $A(\omega)$ as $\omega\to\infty$.

We use the SU(3) anisotropic lattice plaquette action\cite{klassen} as
 \begin{equation}
   S_{\rm G}
 =
   {\beta_{\rm lat}\over N_c \gamma_{\rm G}}
   \sum_{s,i<j\le  3}\mbox{Re}\mbox{Tr}
   \left\{  1  -  P_{ij}(s)  \right\}
   +
   {\beta_{\rm lat} \gamma_{\rm G} \over N_c}
   \sum_{s, i   \le  3}\mbox{Re}\mbox{Tr}
   \left\{ 1  - P_{i4}(s)\right\},
 \end{equation}
where $P_{\mu\nu}(s) \in  \mbox{SU}(3)$ denotes the plaquette operator
in the $\mu$-$\nu$-plane.
The lattice parameter and the  bare anisotropic parameter are fixed as
$\beta_{\rm  lat}  \equiv  2N_c/g^2  =  6.25$ and  $\gamma_{\rm  G}  =
3.2552$, respectively, so as  to reproduce the renormalized anisotropy
as $a_s/a_t = 4$.  The scale  unit is introduced from the on-axis data
of  the   static  inter-quark   potential  with  the   string  tension
$\sqrt{\sigma} = 440$ MeV. The resulting lattice spacings are given as
$a_t^{-1} = 9.365(66)$ GeV  and $a_s^{-1}=2.341(16)$ GeV. The critical
temperature   is  estimated  as   $T_c  \simeq   280$  MeV   from  the
susceptibility   of  the   Polyakov  loop.\cite{ishii2}   We  generate
5,500--9,900   gauge   configurations   to  construct   the   glueball
correlators,\cite{ishii,ishii2}  where  statistical  data are  divided
into bins  of the size  100 to reduce possible  auto-correlations near
$T_c$.  To  enhance the low-energy  spectrum, we adopt  an appropriate
smearing corresponding to $\rho \simeq 0.4$ fm in \Eq{smearing}.

\begin{figure}[ht]
\vspace{-0.5cm}
\centerline{
\includegraphics[angle=-90,scale=0.28]{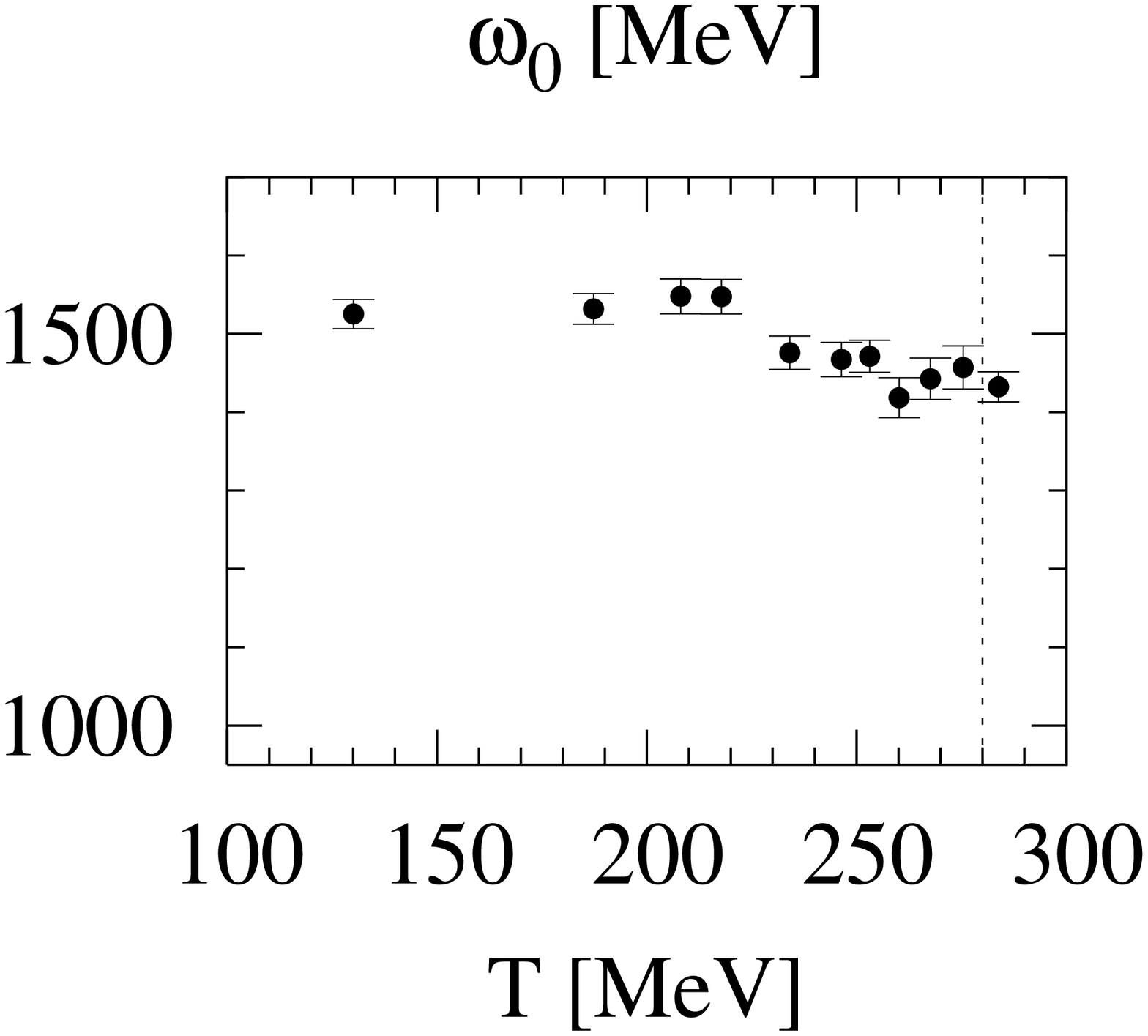}\Bs\Bs\Bs
\includegraphics[angle=-90,scale=0.28]{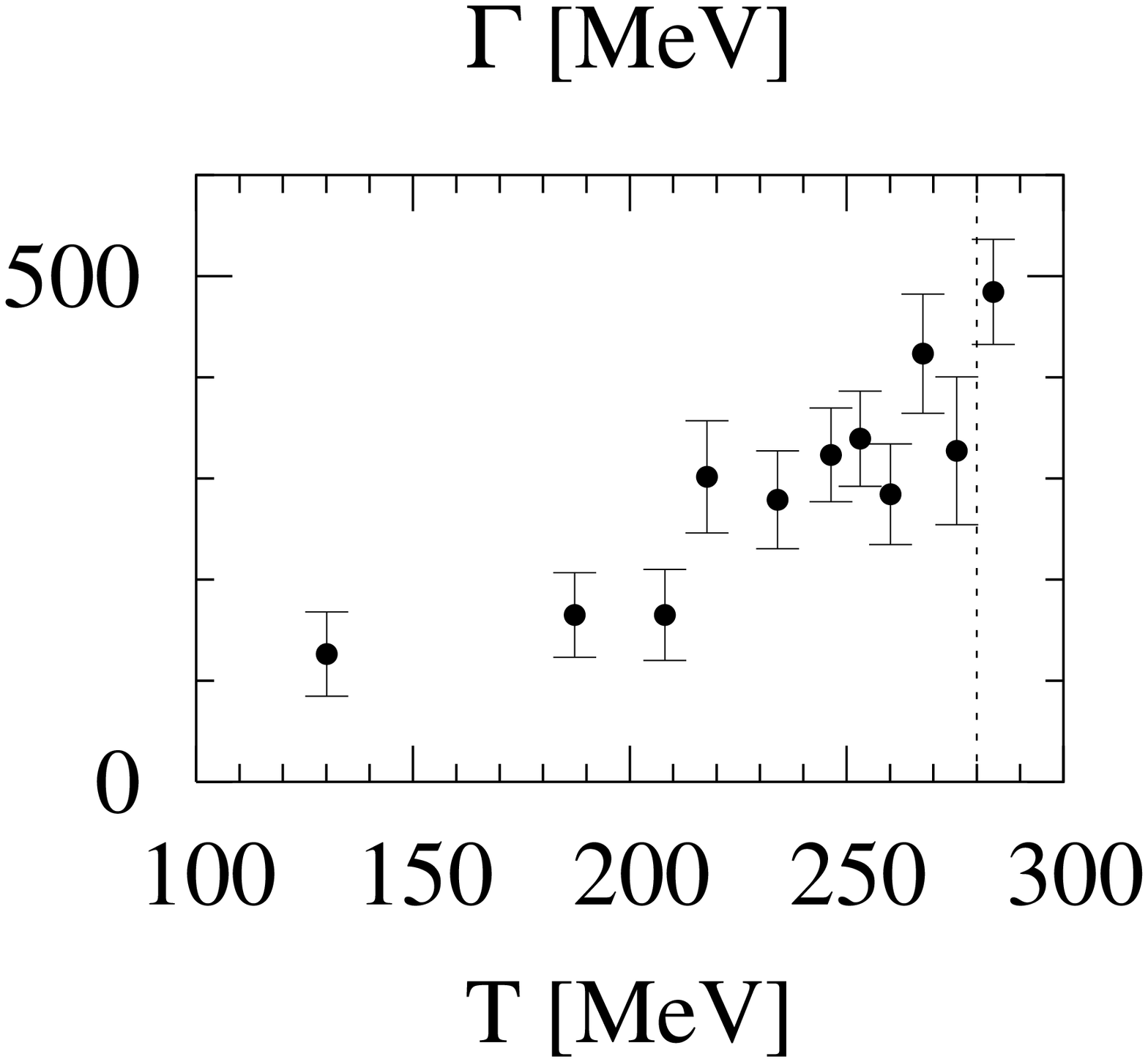}
}
\caption{The   peak  center  $\omega_0(T)$   and  the   thermal  width
$\Gamma(T)$ of the $0^{++}$ glueball against temperature $T$, obtained
from  the $\chi^2$  fit analysis  with the  Breit-Wigner  Ansatz.  The
vertical dotted  lines indicate  the critical temperature  $T_c \simeq
280$ MeV. \label{fig.fit}}
\end{figure}

We present  the numerical  results.  Whereas the  narrow-peak analysis
suggests  that the  thermal effect  on the  glueball is  the pole-mass
reduction\cite{ishii}  of  about 300  MeV,  the advanced  Breit-Wigner
analysis suggests  that it is  a thermal-width broadening\cite{ishii2}
of about 300  MeV together with a modest reduction  in the peak center
of about 100 MeV.  The  numerical results of the $\chi^2$ fit analyses
with the Breit-Wigner Ansatz are shown in \Fig{fig.fit}.

\begin{figure}[ht]
\vspace{-0.3cm}
\centerline{\includegraphics[angle=-90,scale=0.28]{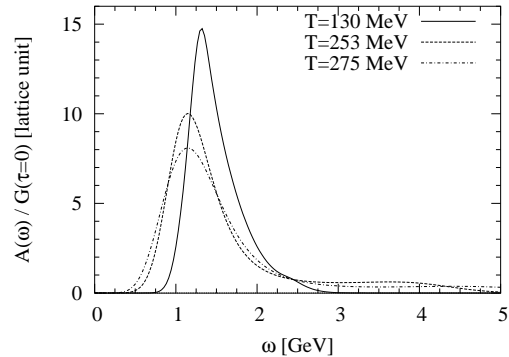}}
\caption{MEM  results for  the $0^{++}$  glueball spectrum  at various
temperatures below $T_c$. \label{fig.mem}}
\end{figure}

We  now turn  to the  result  of MEM  analysis. As  the default  model
function,  we  adopted  the  perturbative expression  of  the  smeared
glueball correlator up to $O(\alpha_{\rm S}^0)$ as\cite{ishii4}
\begin{equation}
  m(\omega)
=
  N\omega^4 \exp\left\{ -(\omega\rho)^2/4 \right\},
\end{equation}
where  the   normalization  factor  $N$   is  determined  from   $1  =
\int_0^{\infty}  d\omega  K(\tau=0,\omega)m(\omega)$  so as  to  mimic
$G(\tau=0)=1$, which we adopt as the normalization of the correlator.
Following the numerical  procedure given in Ref.~\refcite{asakawa}, we
reconstruct $A(\omega)$. The results at various temperatures are shown
in \Fig{fig.mem}.  We  see that the peak tends  to become broader with
the increasing temperature,\cite{ishii4}  which is consistent with the
results of the Breit-Wigner fit analysis.


To  summarize, we  have studied  the properties  of  thermal glueballs
below $T_c$  by using  SU(3) anisotropic lattice  QCD at  the quenched
level.
We have performed $\chi^2$ fit analyses adopting two Ans\"atze for the
spectral  function  $A(\omega)$,  i.e., the  conventional  narrow-peak
Ansatz ($\delta$-function  type) and the  advanced Breit-Wigner Ansatz
so as to respect the appearance of the thermal width at $T>0$.
Whereas  the  former  has  suggested  that  a  significant  pole  mass
reduction of  the $0^{++}$ glueball of  about 300 MeV  near $T_c$, the
latter  has suggested  that the  thermal  effects on  the glueball  is
rather  a  significant  thermal-width  broadening  of  about  300  MeV
together with a modest reduction in peak center of about 100 MeV.
We have  also performed MEM analysis.   We have seen that  the peak of
the  $0^{++}$ glueball  tends to  become broader  with  the increasing
temperature below  $T_c$. This is  consistent with the results  of the
advanced Breit-Wigner  fit analysis, suggesting the  importance of the
thermal width.
Experimentally,  there are  two glueball  candidates,\cite{seth} i.e.,
$f_0(1500)$ and  $f_0(1710)$, both of  which have rather  narrow decay
width $\Gamma \simeq 100$ MeV.  Hence, the experimental observation of
thermal-width broadening  of 300 MeV  would be possible in  the future
experiments at RHIC.  For  more realistic comparison with experiments,
it is desirable to estimate the effects of the dynamical quarks on the
thermal properties of the glueball.


\end{document}